\documentclass[aps,prl,twocolumn,groupedaddress,showpacs]{revtex4}
\usepackage{graphicx,amsmath}
\usepackage{amssymb}

\begin{document}
\title{Dynamical cluster approximation study of the anisotropic two-orbital Hubbard model}

\author{Hunpyo Lee}
\author{Yu-Zhong Zhang}
\author{Harald O. Jeschke}
\author{Roser Valent\'\i}
\affiliation{Institut f\"ur Theoretische Physik, Goethe-Universit\"at
Frankfurt, Max-von-Laue-Stra{\ss}e 1, 60438 Frankfurt am Main, Germany}
\author{Hartmut Monien}
\affiliation{Bethe Center for Theoretical Physics, Universit\"at Bonn, 53115 Bonn, Germany}
\date{\today}

\begin{abstract}

  We investigate the properties of a two-orbital Hubbard model with
  unequal bandwidths on the square lattice in the framework of the
  dynamical cluster approximation (DCA) combined with a
  continuous-time quantum Monte Carlo (CT QMC) algorithm. We explore
  the effect of short-range spatial fluctuations on the nature of the
  metal-insulator transition and the possible occurrence of an
  orbital-selective Mott transition (OSMT) as a function of cluster
  size $N_c$.  We observe that for $N_c=2$ no OSMT is present, instead
  a band insulator state for both orbitals is stabilized at low
  temperatures due to the appearance of an artificial local ordered
  state. For $N_c=4$ the DCA calculations suggest the presence of five
  different phases which originate out of the cooperation and
  competition between spatial fluctuations and orbitals of different
  bandwidths and an OSMT phase is stabilized. Based on our results, we
  discuss the nature of the gap opening.

\end{abstract}

\maketitle

The correlation driven metal-insulator transition (MIT) in
two-dimensional (2D) correlated systems is still poorly
understood. While the behavior of 2D one-band systems at half filling
seems to be settled, this is not the case for multi-orbital systems.
In one-band systems long range correlations or local order in small
size clusters caused by perfect nesting (Slater
physics)~\cite{Slater51,Moukouri01,Kyung03,Gull08} open a gap in the
weak-coupling regime, whereas in the strong-coupling regime the
on-site Coulomb repulsion is the driving force for the gap opening
(Mott physics)~\cite{Mott49}. On the other hand, even though
multi-orbital models are better suited to describe real materials~\cite{Imada98} and
display rich phase diagrams~\cite{Rozenberg97,Florens02,Inaba05} as
well as interesting physics like the orbital-selective Mott transition
(OSMT)~\cite{Anisimov02,Fang04}, they are still under debate.  The
reason for that is their more complex structure compared to the
one-band model due the orbital degrees of freedom, crystal field
splitting effects and the Hund's rule exchange coupling. This
situation gives us a strong motivation to investigate the two-orbital
system.

The OSMT has recently been intensively studied in the context of a
weakly correlated band coexisting and interacting with a more strongly
correlated one in a two-orbital
system~\cite{Knecht05,Medici05,Arita05,Ferrero05,Koga04,Werner07,Liebsch03,Costi07,Bouadim09}.
Issues like (1) the importance of full or Ising-type Hund's rule
coupling~\cite{Knecht05,Arita05,Koga04,Liebsch03}, (2) the
consequences of anisotropic Hund's rule coupling~\cite{Costi07}, (3)
the role of the ratio of the two bandwidths~\cite{Ferrero05}, (4) the
inclusion of the hybridization between bands~\cite{Medici05}, (5) the
effect of crystal field splitting~\cite{Werner07} and (6) the
extension to the three-band case~\cite{Medici09} have
already been addressed. Nevertheless, the importance of spatial
fluctuations has not yet been explored since most calculations have
been performed within the single-site dynamical mean field theory
(DMFT)~\cite{Metzner89,Georges96} where spatial fluctuations are
completely ignored. On the other hand, it has already been noticed
that even in a single-band case, inclusion of spatial correlation will
qualitatively change the scenario of the Mott
MIT~\cite{Gull08,Zhang07,Park08}. Therefore, it is crucial to address 
the effect of spatial fluctuations on the OSMT and the phase diagram.

Very recently, Bouadim {\it et al.}~\cite{Bouadim09} studied the OSMT
by means of a determinant quantum Monte Carlo method (DQMC) on the
square lattice and showed that an itinerant band can coexist with a
fully localized band in a two-orbital Hubbard model as long as long
range antiferromagnetic correlation is absent. However, since the DQMC
calculation was based on a simplified model where one of the two
orbitals is constrained to be fully localized, it still remains
unclear whether the OSMT survives in the system with spatial
fluctuations or not. Moreover, since previous
DMFT~\cite{Knecht05,Medici05,Arita05,Ferrero05,Koga04,Werner07,Liebsch03,Costi07}
and a slave spin mean field calculation~\cite{Medici09} are based on
the Bethe lattice, it is interesting to move in the direction of real
systems by studying the case of a two-dimensional model on the square
lattice with the Fermi level at a van Hove singularity at half
filling.

In this Letter we concentrate on the nature of the gap opening and the
OSMT in a two-dimensional system. The anisotropic two-orbital Hubbard
model on the square lattice at half-filling is the simplest model
which can describe the OSMT including spatial fluctuations. The
Hamiltonian is given as
\begin{eqnarray*}
  H &=&-\sum_{\langle ij\rangle m\sigma}  t_m c^{\dagger}_{im\sigma}
  c_{jm\sigma}\,+\,U\sum_{im}n_{im\uparrow} n_{im\downarrow}\\
    &&+\sum\nolimits_{i\sigma\sigma'}(U'-\delta_{\sigma \sigma'} J_z)n_{i1\sigma}
  n_{i2\sigma'},
\end{eqnarray*}
where $t_m$ for orbital $m=(1,2)$ denotes the hopping integrals
between nearest-neighbor (n.n.) sites $i$ and $j$, $U$ and $U'$ are
intra-orbital and inter-orbital Coulomb repulsion integrals
respectively and $J_z n_{i1 \sigma}n_{i2 \sigma}$ for spin $\sigma$ is
the Ising-type Hund's rule coupling term. In our calculations we
ignore spin-flip and pair-hopping processes. We also set $t_1/t = 0.5$
(narrow band), $t_2/t = 1.0$ (wide band), $J_z=U/4$ and $U'=U/2$. For
this model we employ the dynamical cluster approximation (DCA) method
with cluster sizes $N_c=2$ and 4. The DCA
method~\cite{Maier05,Hettler98,Hettler00} can not only overcome the
problem of the single-site DMFT method~\cite{Metzner89,Georges96},
where Mott physics rather than Slater physics is emphasized in the
paramagnetic phase due to the lack of spatial fluctuations, but it is
also computationally cheaper than lattice calculations. We use a weak
coupling continuous-time quantum Monte Carlo algorithm as an impurity
solver~\cite{Rubtsov05,Lee08}. We shall present results on the
spin-spin correlations, double occupancy, self-energy and density of
states (DOS).

\begin{figure}
\includegraphics[angle=-90,width=0.4\textwidth]{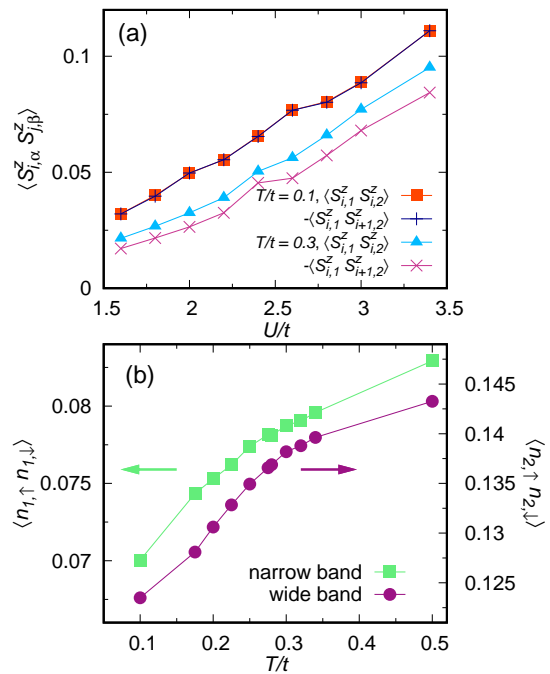}
\caption {\label{Nc2_result} (a) On-site and inter-site inter-orbital
  spin-spin correlations for $N_c=2$ as a function of $U/t$ at
  temperatures $T/t=0.1$ and $T/t=0.3$. (b) Double occupancy for
  $N_c=2$ as a function of $T/t$ for $U/t=2.4$. Left and right axes
  are for the double occupancy of the narrow band and the wide band,
  respectively.}
\end{figure}

First, let us discuss the results obtained from the DCA with a
two-site cluster ($N_c=2$). It is known that, for a one-band system at
low temperatures, the formation of a local singlet state driven by
Slater physics is responsible for the gap opening. In the two-band
system orbital fluctuations are present. Due to the Hund's rule
coupling and the Coulomb interaction, ferromagnetic (FM) correlation
between orbitals and antiferromagnetic (AF) correlation between sites
develop. In order to check for these correlations we measure the
on-site (inter-site) inter-orbital spin-spin correlations $\langle
S_{i,1}^z S_{i,2}^z\rangle$ ($\langle S_{i,1}^z S_{i+1,2}^z\rangle$)
as a function of $U/t$. The results are shown in
Fig.~\ref{Nc2_result}(a).

As the Hund's rule coupling $J_z= U/4$ is increased, the on-site
inter-orbital FM correlations (positive sign) and inter-site
inter-orbital AF correlations (negative sign) are enhanced for both
temperatures $T/t=0.3$ and $T/t=0.1$. At high temperatures $T/t=0.3$
the on-site inter-orbital FM correlations are stronger than the
inter-site inter-orbital AF correlations which can be attributed to
the fact that thermal fluctuations suppress the AF correlations. At
$T/t=0.1$ both correlations are of the same magnitude. This behavior
suggests the appearance of a local ordered state in the low
temperature regime.  In order to verify whether this state, in analogy
to the one-band model, is responsible for the gap opening as described
by Slater physics, we calculate the temperature dependence of double
occupancy for both orbitals. If Slater physics is dominant, as the
temperature is decreased the formation of the local order which
reduces the potential energy $U\langle
n_{\uparrow}n_{\downarrow}\rangle$ should cause the gap opening. In
Fig.~\ref{Nc2_result}(b) we present the double occupancy as a function
of temperature $T/t$ for $U/t=2.4$. The double occupancy in both bands
decreases with decreasing temperature and it shows a more abrupt drop
near $T/t=0.2$. This behavior gives strong evidence of Slater physics,
and the band insulator in both orbitals should be present at zero
temperature for all positive interaction strengths $U/t$.

\begin{figure}
\includegraphics[angle=-90,width=0.4\textwidth]{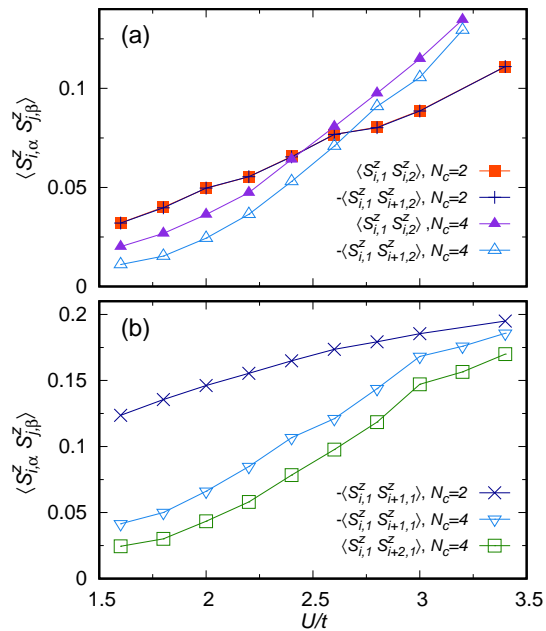}
\caption {\label{Nc4_spincorrelation} (a) The on-site (inter-site)
  inter-orbital spin-spin correlations for $N_c=2$ and 4 as a function
  of $U/t$ at $T/t=0.1$. (b) The narrow band (next) nearest-neighbor
  spin-spin (n.n.n.,n.n.) correlations for $N_c=2$ and 4 as a function of
  $U/t$ at $T/t=0.1$}
\end{figure}

Next, we explore the DCA for a 4-site cluster ($N_c=4$). The inclusion
of next nearest-neighbor (n.n.n.) correlations in $N_c=4$ suppresses the
local ordered state enhanced artificially in the $N_c=2$ cluster. In
addition, the system shows a weak degree of frustration because of the
absence of long range correlations. In this way, the Mott physics
present in the single-site systems coexists
with Slater physics present in the two-site systems. Therefore we believe that the description in terms of
the $N_c=4$ clusters is closer to the real materials at finite
temperature.  In Fig.~\ref{Nc4_spincorrelation}(a) we compare the
on-site and inter-site inter-orbital spin-spin correlation results for
$N_c=2$ and 4 for $T/t=0.1$.  The same magnitude of both correlations
for $N_c=2$ implies the presence of a relatively strong inter-site
local ordered state, while the deviation of those for $N_c=4$
indicates that the local ordered state is released due to the influence
of the n.n.n. correlations.  In order to compare directly the strength of
this local ordered state for $N_c=2$ and 4, we plot n.n. and n.n.n.
correlations for the narrow band in Fig.~\ref{Nc4_spincorrelation}(b).
In the weak coupling regime the n.n. correlation strength for $N_c=2$
due to enhanced Slater physics is larger than that for $N_c=4$.  In
the strong coupling regime the n.n. correlation strength is similar for
$N_c=2$ and 4 since the insulating state for $N_c= 4$ is induced by
cooperation of Mott and Slater physics.  We also find strong n.n.n.
correlations for $N_c=4$.

\begin{figure}
\includegraphics[angle=-90,width=0.4\textwidth]{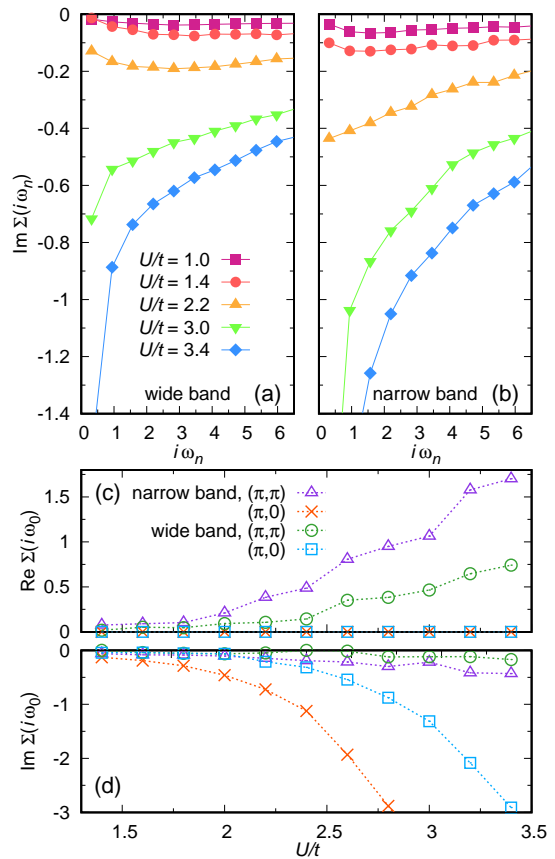}
\caption {\label{Nc4_selfenergy} The imaginary part of the on-site
  self-energy for $U/t=1.0$,1.4,2.2,3.0,3.4 at $T/t=0.1$ for $N_c=4$
  (a) in the narrow band and (b) in the wide band. The real (c) and
  imaginary (d) part of self-energy at the lowest Matsubara frequency
  $\omega_0$ for $K$=($\pi$,$\pi$) and ($\pi$,0) sectors as a function
  of $U/t$.}
\end{figure}

The competition among magnetic and orbital fluctuations as well as
weak frustration for $N_c = 4$ should generate a rich phase
diagram. In order to investigate this complex situation we analyze, in
what follows, the on-site self-energy. The imaginary part of the
on-site self-energy $\rm{Im}\,\Sigma (i\omega_n)$ provides information
about the possible Fermi-liquid/non-Fermi-liquid behavior of the
system as well as the nature of the gap opening. In
Figs.~\ref{Nc4_selfenergy}(a) and (b) we present $\rm{Im} \,\Sigma
(i\omega_n)$ for the narrow and wide band, respectively, at
$T/t=0.1$. According to Fermi-liquid theory, $\rm{Im}\,\Sigma
(\omega)$ at $T=0$ at $\omega \rightarrow 0$ extrapolates to 0. In the
weak-coupling regions below $U/t=1.4$ this Fermi-liquid behavior is
seen in both bands. Between $U/t=1.4$ and 1.8 Fermi-liquid behavior is
still present in the wide band, while non-Fermi-liquid behavior is
observed in the narrow band. The electrons begin to localize in the
narrow band driven by both Slater and Mott physics, while those in the
wide band are still delocalized. As the interaction is increased,
non-Fermi-liquid behavior is observed in both bands. At $U/t=2.8$,
$\rm{Im}\,\Sigma (i\omega_n)$ in the narrow band diverges, which
indicates the opening of a gap, while the
metallic state (non-Fermi-liquid) is still present in the wide
band. These results evidence a OSMT. In the strong-coupling region
$U/t=3.4$ the insulating state is observed in both bands.

\begin{figure}
\includegraphics[angle=-90,width=0.4\textwidth]{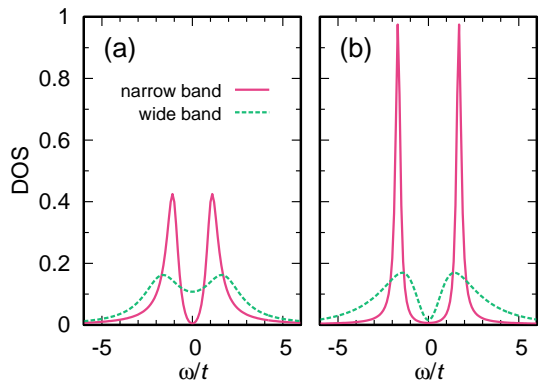}
\caption {\label{Nc4_DOS} Density of states at $T/t=0.1$ and $N_c=4$
for (a)
  $U/t=2.8$ and (b) $U/t=3.4$. We employ the Pad{\'e}
  approximation method for the analytic continuation.}
\end{figure}

\begin{figure}
\includegraphics[width=0.4\textwidth]{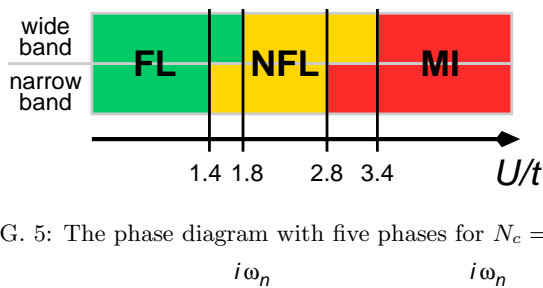}
\caption {\label{phases} The phase diagram with five phases for $N_c=4$.}
\end{figure}

In what follows we shall analyze the nature of the gap opening.
According to recent results obtained for the single-band plaquette
Hubbard model~\cite{Park08}, momentum sectors
$K$=$(0,0)$/$(\pi,\pi)$ and $(\pi,0)$/$(0,\pi)$ undergo a metal to
band insulator transition and a metal to Mott insulator transition,
respectively.
In Figs.~\ref{Nc4_selfenergy} (c) and (d), respectively, we present
the real and imaginary parts of the self-energy at the lowest Matsubara
frequency $\omega_0$, $\rm{Re}\,\Sigma(i\omega_0)$ and
$\rm{Im}\,\Sigma(i\omega_0)$, for $K$=($\pi$,$\pi$) and ($\pi$,0) in
both bands.  While $\rm{Re}\,\Sigma(i\omega_0)$ gives information
about the energy shift of the spectral function,
$\rm{Im}\,\Sigma(i\omega_0)$ introduces the scattering rate. As the
interaction is increased, $\rm{Re}\,\Sigma_{ (\pi ,
  \pi)}=-\rm{Re}\,\Sigma _{(0,0)}$ increases while
$\rm{Im}\,\Sigma_{(\pi , \pi)}=\rm{Im}\,\Sigma_{ (0 , 0)}$ remains
small in both bands. These results suggest a metal to band insulator
transition where the gap is opened through separation of the poles
away from the Fermi level. On the other hand, as the interaction
increases $\rm{Im}\,\Sigma_{ (\pi , 0)}=\rm{Im}\,\Sigma_{ (0 , \pi)}$
displays a divergent behaviour and $\rm{Re}\,\Sigma_{ (\pi ,
  0)}=-\rm{Re}\,\Sigma_ {(0 , \pi)}$ in both bands is zero due to
particle-hole symmetry. Therefore, in the strong coupling region, the
gap in the $K$=$(\pi ,0)$ and $(0, \pi)$ sectors is only induced by
the divergence of $\rm{Im}\,\Sigma(i\omega_0)$ which is a signature
for Mott physics. These results are similar to the single-band
plaquette Hubbard model results~\cite{Park08} but, while a first-order transition
occurs in the single-band Hubbard model, the OSMT behaviour is present
in the two-band Hubbard model. In order to show the OSMT more clearly,
we present in Fig.~\ref{Nc4_DOS} the density of states (DOS) at
$T/t=0.1$ for the interaction values $U/t=2.8$ (Fig.~\ref{Nc4_DOS}
(a)) and $U/t=3.4$ (Fig.~\ref{Nc4_DOS} (b)). These interaction values
have been identified as onsets for the OSMT and insulator phases,
respectively. For $U/t=2.8$ the narrow band exhibits a gap at the
Fermi energy ($\omega=0$), while the wide band has a finite DOS at
$\omega=0$.
This means that for a given interaction strength $U/t$ two stages of
the Mott transition are present, with a Mott insulator in the narrow
band and a metal in the wide band. At $U/t=3.4$ both bands show a
gap at $\omega=0$. The gap in the narrow band is wider than that in
the wide band. Finally, we plot the phase diagram with the identified
five phases for $N_c=4$ in Fig.~\ref{phases}.

In summary, we have explored the anisotropic two-orbital Hubbard
model using the DCA method with cluster sizes $N_c=2$ and 4. The DCA
cluster with $N_c=2$ for the single-band model is known to describe
a system with artificially strong local order between sites and the
gap opening is controlled by Slater physics. Our results show
that this inter-site AF correlation is still strong in spite of
orbital fluctuations, leading to a gap at low
temperatures. The appearance of the insulating states can be
described by Slater physics. We have also investigated within DCA
the $N_c=4$ cluster which includes n.n.n. correlations. Unlike the
$N_c=2$ cluster, the  local ordered states are not present
in the weak-coupling limit. 
 In the very weak-coupling
 regime Fermi-liquid behavior is present in both bands.
 As the interaction increases, the electrons in the narrow band weakly
localize and non-Fermi-liquid behavior is observed, even though the
Fermi-liquid behavior is still present in the wide band. In the
intermediate region, non-Fermi-liquid behavior is observed in both
bands. In the strong-coupling region the electrons in the narrow
band are completely localized and those in the wide band are
partially localized which can be described as the OSMT. In the very
strong-coupling region both orbitals are insulating. The nature of
the  gap opening is that of coexisting Slater physics in the
momentum sector $K$=$(0,0)$/$(\pi,\pi)$ and Mott physics in momentum
sector $K$=$(\pi,0)$/$(0,\pi)$.

{\it Acknowledgments.-} We thank N. Bl\"umer and A. M. Ol{\'e}s for
useful comments and the Deutsche Forschungsgemeinschaft for financial
support through the SFB/TRR~49 and Emmy Noether programs, and we
acknowledge support by the Frankfurt Center for Scientific Computing.
 
\bibliography{paper8}

\end{document}